\documentclass[aps,prl,twocolumn, superscriptaddress,amsmath,amssymb, notitlepage,reprint,footinbib]{revtex4-1}
\usepackage{graphicx}
\usepackage{amsmath}
\usepackage{amsfonts}
\usepackage{amssymb}
\usepackage{booktabs}
\usepackage{hyperref}
\usepackage{xcolor}
\usepackage{blkarray}

\hypersetup{
    colorlinks,
    linkcolor={red!50!black},
    citecolor={red!50!black},
    urlcolor={red!50!black}
}
\newcommand{\MYhref}[3][blue]{\href{#2}{\color{#1}{#3}}}%

\begin{document}
\title{Tight-Binding Studio: A Technical Software Package to Find the Parameters of Tight-Binding Hamiltonian}

\affiliation{Department of Physics, University of Antwerp, Groenenborgerlaan 171, B-2020 Antwerp, Belgium.}
\affiliation{School of Physics, University of Damghan, P.O. Box 36716-41167, Damghan, Iran.}

\author{M. \textsc{Nakhaee}$^{1,2}$}
\email{mohammad.nakhaee@uantwerpen.be}
\author{S. A. \textsc{Ketabi}$^{2}$}
\email{saketabi@du.ac.ir}
\author{F. M. \textsc{Peeters}$^{1}$}
\email{francois.peeters@uantwerpen.be}

\date{\today }

\begin{abstract}
We present Tight-Binding Studio (TBStudio) software package for calculating tight-binding Hamiltonian from a set of Bloch energy bands obtained from first principle theories such as density functional theory, Hartree-Fock calculations or Semi-empirical band structure theory. This will be helpful for scientists who are interested in studying electronic properties of structures using Green's function theory in tight-binding approximation. TBStudio is a cross-platform application written in C++ with a graphical user interface design that is user-friendly and easy to work with. This software is powered by Linear Algebra Package C interface library for solving the eigenvalue problems and the standard high performance OpenGL graphic library for real time plotting. TBStudio and its examples together with the tutorials are available for download from \MYhref{https://tight-binding.com}{\textit{tight-binding.com}}
\end{abstract}

\maketitle


\section{Introduction}
\label{sec:introduction}
Many researchers are interested in the study of nanostructures and solid state materials in general, but there are many computational and mathematical challenges which hinder rapid progress. Therefore, new computational packages are urgently needed in order to accelerate scientific progress. Here, we are interested in electronic structure properties. 

First-principles electronic structure calculations \cite{FP} are based on the laws of quantum mechanics and only use the fundamental constants of physics as input to provide detailed insight into the origin of mechanical, electronic, optical and magnetic properties of materials and molecules. Both structural and spectroscopic information is directly obtainable from high-performance computations and yields information which is complementary to that obtained from experimental such as transmission electron microscopy data.

Density functional theory (DFT) is one of the most important methods used in electronic structure calculations and computational physics has provided already a diverse number of software packages such as OpenMX \cite{openmx}, VASP \cite{VASP1, VASP2, VASP3}, QUANTUM ESPRESSO \cite{QE} and ABINIT \cite{ABINIT} that are based on different algorithms. The accuracy of different methods depends on the used approximations, e.g. the particular form for the exchange-correlation (XC) energy.

First-principles calculations is commonly applied to calculate the properties of an infinite periodic arrangement of one or more atoms (the basis) repeated at each lattice point that describes a highly ordered structure, occurring due to the intrinsic nature of its constituents to form symmetric patterns. The periodicity can make problems easier, but sometimes it can be a drawback when one is interested in the electronic properties of finite size systems. Linear combination of atomic orbitals (LCAO) \cite{lcao1,lcao2} is a good candidate to overcome this problem.

The most important justification to setup LCAO is that the combination of this method with Green's function
theory can be also used for non-periodic systems and furthermore, in the case of systems with a huge number of atoms
there are a variety of cost and time efficiency motivations which can lead one to use the LCAO method.

Tight-Binding (TB) approaches are based on the LCAO method that is primarily used to calculate the band structure and single-particle Bloch states of a material as e.g. done by Slater and Koster (SK) \cite{slaterkoster}. The tight-binding method is simple and computationally very fast. Therefore, it is often used in calculations of very large systems, with more than a few thousand atoms in the unit cell. There are a number of earlier reviews ~\cite{Ackland,Goringedag} that people working in this field should be aware of. TB Hamiltonian gives the electronic structure of a system using a real-space picture of the system. The real Hamiltonians provide insight into the nature of the transport mechanisms.

To find a TB Hamiltonian one needs to evaluate the band-structure of a typical structure based on first principles calculations and construct a tight-binding model via the Slater and Koster method to reproduce the band energies obtained from DFT.

The purpose of this paper is to introduce Tight-Binding Studio (TBStudio), a new technical software package for generating TB Hamiltonians based on Slater-Koster method from data obtained from first principles calculations that has been made available at \MYhref{https://tight-binding.com}{\textit{tight-binding.com}}. Cross-platform graphical user interface of TBStudio is written in C++ using native controls and emulates foreign functionality via wxWidgets \cite{wxWid} tools library for GTK, MS Windows, and MacOS. Also, BLAS \cite{BLAS} and LAPACK \cite{LAPACK} routines are used for matrix multiplications, solving systems of simultaneous linear equations and eigenvalue problems. The standard high performance OpenGL graphic library \cite{openglsite} has been used for real time plotting. The main structure of TBStudio and several important topics, including the post processing tools are explained in the rest of this paper.

\section{Linear Combination of Atomic Orbitals}
\label{sec:SK}
Consider two atoms which have atomic orbitals described by wave function $\Psi_A$ and $\Psi_B$. If the atoms are at the equilibrium distance, their electron clouds overlap with each other and the wave function of molecular orbital can be obtained from a linear combination of atomic orbitals as follows

\begin{align}
\Psi_{AB} = c_A \Psi_A + c_B \Psi_B \, ,
\label{eqn:psi}
\end{align}

\noindent where $\Psi_{AB}$ is the normalized wave function of molecular orbitals of the molecule AB. With this in mind we are able to expand the Bloch functions as linear combinations of the orbitals $\varphi$ as follows

\begin{equation}
\centering
\psi_k(\textbf{r}) = \sum\limits_{i} \sum\limits_{\nu_i} c_{i \nu_i}(k) \phi_{\nu_i,k}(\textbf{r}-\textbf{r}_i) \, ,
\label{eqn:Bloch}
\end{equation}
\noindent where $i$ runs over all unit cell atoms and $\nu_i$ runs over the orbitals defined for $i^{th}$ atom and
\begin{align}
\centering
\phi_{\nu,\textbf{k}}(\textbf{r}) = \sum\limits_{n \in \mathbb{Z}} \sum\limits_{m \in \mathbb{Z}} \sum\limits_{l \in \mathbb{Z}} e^{i \textbf{k}.\textbf{R}_{n,m,l}} \varphi_{\nu}(\textbf{r}-\textbf{R}_{n,m,l}) \, ,
\label{eqn:lcao}
\end{align}

\noindent in which $\textbf{R}_{n,m,l}$ is the discrete translation vector of the unit cell at $(n,m,l)$ of the Bravais lattice. The electron hopping between different orbitals is an essential ingredient in TB approach. In a simple non-interacting picture, the overlap of the outermost electrons leads to the hybridization of the electronic orbitals and leads to the de-localization of Bloch states.

To calculate the energy bands we should solve the generalized eigenvalue problem

\begin{align}
\centering
\hat{H}(\textbf{k}) \psi_n^{\sigma}(\textbf{k})= \varepsilon_{n \boldsymbol k \sigma} \hat{S}(\textbf{k}) \psi_n^{\sigma}(\textbf{k}) \, ,
\label{eqn:EVP}
\end{align}

\noindent where $\hat{H}$ and $\hat{S}$ are the TB Hamiltonian and the overlap operators that can be written as

\begin{align}
\centering
\hat{H} &= \sum\limits_{i,i'} \sum\limits_{\nu_i,\nu_i'} \hat{H} |\phi_{\nu_i}><\phi_{{\nu'}_{i'}}| \, , \nonumber \\
\hat{S} &= \sum\limits_{i,i'} \sum\limits_{\nu_i,\nu_i'} |\phi_{\nu_i}><\phi_{{\nu'}_{i'}}| \, ,
\label{eqn:SQTBHS}
\end{align}

\noindent where in general the basis can be non-orthogonal and then the overlap matrix can be a non-identity matrix. The elements of the Hamiltonian and the overlap matrices can be found by definition of the molecular two-center integrals in terms of the quantities called Slater and Koster integrals.

\section{Calculation of Slater-Koster integral Table}
\label{sec:SK}
The most important issue that is the background of the idea of Slater and Koster approach is the rotation operator defined as follows

\begin{align}
{\displaystyle D(\mathbf {\hat {n}} ,\phi )=\exp \left(-i\phi {\frac {\mathbf {\hat {n}} \cdot \mathbf {J} }{\hbar }}\right)} \, ,
\label{eqn:D}
\end{align}

\noindent and the SK parameters

\begin{align}
h_{l l'}^{m m'}(\mathbf{r}) &= \left\langle \varphi_l^m(\mathbf{x} + \mathbf{r}) | H(\mathbf{x} + \mathbf{r}) | \varphi_{l'}^{m'}(\mathbf{x}) \right\rangle \nonumber \, ,
\\
s_{l l'}^{m m'}(\mathbf{r}) &= \left\langle \varphi_l^m(\mathbf{x} + \mathbf{r}) | \varphi_{l'}^{m'}(\mathbf{x}) \right\rangle  \, ,
\label{eqn:SK}
\end{align}

\noindent in which $\varphi_{l}^{m}(\mathbf{x})$ stands for a specific atomic orbital defined by angular quantum numbers $l$ and $m$. Note,
the orbitals in Eq. \eqref{eqn:SK} are the real spherical harmonics. Without any change in the basic
framework one can rotate the basis vectors as follows

\begin{align}
h_{l l'}^{m m'}(\mathbf{r}) &= \left\langle \varphi_l^m(\mathbf{x} + \mathbf{r}) | \bar{D}^\dagger H(\mathbf{x} + \mathbf{r}) \bar{D} | \varphi_{l'}^{m'}(\mathbf{x}) \right\rangle \nonumber \, ,
\\
s_{l l'}^{m m'}(\mathbf{r}) &= \left\langle \varphi_l^m(\mathbf{x} + \mathbf{r}) |\bar{D}^\dagger \bar{D}| \varphi_{l'}^{m'}(\mathbf{x}) \right\rangle   \, ,
\label{eqn:SK2}
\end{align}

\noindent where $\bar{D}=\bar{D}(l,m,n)$ operator is a function of directional cosines defined by the angles of the bond vector between the atoms and the Cartesian basis vectors $\hat{x}$, $\hat{y}$ and $\hat{z}$. The rotation operator for different orbitals can be calculated from the Clebsch-Gordan coefficients. The bar symbol means symmetrized coefficients which are essential to rotate a real spherical harmonic. The complex spherical harmonics will be converted to real spherical harmonics by applying the following operators ($R_l$) for the different orbitals

\begin{align}
R_0 &= \left(
\begin{array}{c}
1 \\
\end{array}
\right) \nonumber \, , \\
R_1 &= \left(
\begin{array}{ccc}
\frac{i}{\sqrt{2}} & 0 & \frac{i}{\sqrt{2}} \\
0 & 1 & 0 \\
\frac{1}{\sqrt{2}} & 0 & -\frac{1}{\sqrt{2}} \\
\end{array}
\right)  \nonumber \, , \\
R_2 &= \left(
\begin{array}{ccccc}
\frac{i}{\sqrt{2}} & 0 & 0 & 0 & -\frac{i}{\sqrt{2}} \\
0 & \frac{i}{\sqrt{2}} & 0 & \frac{i}{\sqrt{2}} & 0 \\
0 & 0 & 1 & 0 & 0 \\
0 & \frac{1}{\sqrt{2}} & 0 & -\frac{1}{\sqrt{2}} & 0 \\
\frac{1}{\sqrt{2}} & 0 & 0 & 0 & \frac{1}{\sqrt{2}} \\
\end{array}
\right) \, .
\label{eqn:R}
\end{align}

The integrals of the rotated vectors in the right hand side of Eq. \eqref{eqn:SK2} are called SK parameters as ${ss\sigma}$, ${sp\sigma}$, ${pp\sigma}$, ${pp\pi}$, ${sd\sigma}$, ${pd\sigma}$, ${pd\pi}$, ${dd\sigma}$, ${dd\pi}$ and ${dd\delta}$ \cite{slaterkoster,lendi} where the first and the second letters are the orbital label as $s$, $p$ and $d$ and the third letter stands for the type of covalent binding which are formed by the overlap of atomic orbitals. ${\sigma}$, ${\pi}$ and ${\delta}$ bonds are related to the relative directions of two typical orbitals.

\section{Atomic Spin-Orbit Coupling}
\label{sec:soc}
Structures including heavy atoms show a considerable spin-orbit effect in their electronic properties \cite{socdelta}. Experimentally, this phenomenon is detectable as a splitting of spectral lines. The addition of spin-orbit coupling (SOC) to study such materials is known as fine structure. There are many structures \cite{soc1,soc2,soc3,soc4,soc5} in which taking SOC into accounts becomes very important in atomistic modeling. This effect is a phenomenon that comes from a relativistic interaction of a particle's spin with its motion inside a potential $\mathbf{V}$ and so the energy level split produced by the SOC is usually of the same order in size to the relativistic corrections to the kinetic energy. SOC fine structure can be added as a term to the Hamiltonian. The atomic spin-orbit interaction may be included in the TB model as

\begin{align}
\hat{H}_{SO} = \frac{1}{2(m_e c)^2} (\mathbf{\nabla} \mathbf{V} \times \mathbf{P}).\mathbf{S} \, ,
\end{align}

\noindent where $\mathbf{P}$ and $\mathbf{S}$ are momentum and spin, respectively. Spin-orbit interactions can be accurately approximated by a local atomic contribution of the form

\begin{align}
\hat{H}_{SO} = \sum_{i}\sum_{\nu} \hat{P}_{i,\nu} \lambda_{\nu} \mathbf{L}_i.\mathbf{S}_i \, ,
\end{align}

\noindent in which $\hat{P}_{i,\nu}$ is the projection operator for angular momentum $\nu$ on site $i$ and $\lambda_{\nu}$ denotes the SOC constant for the angular momentum $\nu$, and $S_i$ is the spin operator
on site $i$. The additional term in the Hamiltonian can be found by calculating the $\mathbf{L}\otimes\mathbf{S}$ operator. The angular and spin operators can be written as follows

\begin{align}
&\mathbf{L} = (L_x,L_y,L_z) \nonumber \, , \\
&\mathbf{S} = (S_x,S_y,S_z) \, ,
\end{align}

\noindent where

\begin{align}
&L_x = \frac{L^{+}+L^{-}}{2} \nonumber \, , \\
&L_y = \frac{L^{+}-L^{-}}{2 i} \, ,
\end{align}

\noindent in which $L^{+,-}$ denote the ladder operators for the orbitals. Similarly, we obtain $S_x$ and $S_y$ in terms of the $S^{+,-}$ ladder spin operators. Note that we need the operator to be compatible with the real spherical harmonics. The real SOC operator can be evaluated as follows

\begin{align}
\bar{R}_l.(\mathbf{L} \otimes \mathbf{S}).\bar{R}_l^{-1} \, ,
\end{align}

\noindent where

\begin{align}
\bar{R}_l=R_l \otimes I_{2\times2} \, .
\end{align}

One should find the final results for the different $s$, $p$ and $d$ orbitals as listed in Table~\ref{tab1}.
\begin{table*}[ht]
	\centering
	\caption{The real spherical harmonics for different angular quantum numbers $l$ and $m$.}
	\label{tab1}
	\begin{tabular}{@{}cccccccccc@{}}
		\toprule
		\hline
		\hline
		l & $0$ & $1$ & $1$ & $1$ & $2$ & $2$ & $2$ & $2$ & $2$\\
		\hline
		\hline
		m & $0$ & $-1$ & $0$ & $1$ & $-2$ & $-1$ & $0$ & $1$ & $2$\\
		\hline
		Real spherical harmonic  \quad  \quad  & $ \quad s \quad $ & $ \quad p_y \quad $ & $ \quad p_z \quad $ & $ \quad p_x \quad $ & $ \quad d_{\text{xy}} \quad $ & $ \quad d_{\text{yz}} \quad $ & $ \quad d_{3 z^2-r^2} \quad $ & $ \quad d_{\text{xz}} \quad $ & $ \quad d_{x^2-y^2} \quad $ \\
		\hline
	\end{tabular}
\end{table*}

\begin{widetext}
	\begin{equation*}
	\centering
	\hat{H}_{s}^{SOC} = \lambda_s \mathbf{L_s}\otimes\mathbf{S} =
	\left(
	\begin{array}{cc}
	0 & 0 \\
	0 & 0 \\
	\end{array}
	\right)  \, ,
	\end{equation*}
	\begin{equation*}
	\centering
	\hat{H}_{p}^{SOC} = \lambda_p \mathbf{L_p}\otimes\mathbf{S} =\lambda_p \quad
	\begin{blockarray}{*{6}{c} l}
	\begin{block}{*{6}{>{$\footnotesize}c<{$}} l}
	$p_y$ & $p_y$ & $p_z$ & $p_z$ & $p_x$ & $p_x$ & $ $\\
	\end{block}
	\begin{block}{(*{6}{c})>{$\footnotesize}l<{$}}
	0 & 0 & 0 & \frac{i}{2} & -\frac{i}{2} & 0 & $\quad p_y$\\
	0 & 0 & \frac{i}{2} & 0 & 0 & \frac{i}{2} & $\quad p_y$\\
	0 & -\frac{i}{2} & 0 & 0 & 0 & \frac{1}{2} & $\quad p_z$\\
	-\frac{i}{2} & 0 & 0 & 0 & -\frac{1}{2} & 0 & $\quad p_z$\\
	\frac{i}{2} & 0 & 0 & -\frac{1}{2} & 0 & 0 & $\quad p_x$\\
	0 & -\frac{i}{2} & \frac{1}{2} & 0 & 0 & 0 & $\quad p_x$\\
	\end{block}
	\end{blockarray}  \, \, \, ,
	\end{equation*}
	\vspace{-0.2cm}
	\begin{equation*}
	\centering
	\hat{H}_{d}^{SOC} = \lambda_d \mathbf{L_d}\otimes\mathbf{S} =\lambda_d \quad
	\begin{blockarray}{*{10}{c} l}
	\begin{block}{*{10}{>{$\footnotesize}c<{$}} l}
	$d_{\text{xy}}$ & $d_{\text{xy}}$ & $d_{\text{yz}}$ & $d_{\text{yz}}$ & $d_{3 z^2-r^2}$ & $d_{3 z^2-r^2}$ & $d_{\text{xz}}$ & $d_{\text{xz}}$ & $d_{x^2-y^2}$ & $d_{x^2-y^2}$ & $ $\\
	\end{block}
	\begin{block}{(*{10}{c})>{$\footnotesize}l<{$}}
	0 & 0 & 0 & -\frac{1}{2} & 0 & 0 & 0 & \frac{i}{2} & -i & 0 & $\quad d_{\text{xy}}$\\
	0 & 0 & \frac{1}{2} & 0 & 0 & 0 & \frac{i}{2} & 0 & 0 & i & $\quad d_{\text{xy}}$\\
	0 & \frac{1}{2} & 0 & 0 & 0 & \frac{i \sqrt{3}}{2} & -\frac{i}{2} & 0 & 0 & \frac{i}{2} & $\quad d_{\text{yz}}$\\
	-\frac{1}{2} & 0 & 0 & 0 & \frac{i \sqrt{3}}{2} & 0 & 0 & \frac{i}{2} & \frac{i}{2} & 0 & $\quad d_{\text{yz}}$\\
	0 & 0 & 0 & -\frac{i \sqrt{3}}{2} & 0 & 0 & 0 & \frac{\sqrt{3}}{2} & 0 & 0 & $\quad d_{3 z^2-r^2}$\\
	0 & 0 & -\frac{i \sqrt{3}}{2} & 0 & 0 & 0 & -\frac{\sqrt{3}}{2} & 0 & 0 & 0 & $\quad d_{3 z^2-r^2}$\\
	0 & -\frac{i}{2} & \frac{i}{2} & 0 & 0 & -\frac{\sqrt{3}}{2} & 0 & 0 & 0 & \frac{1}{2} & $\quad d_{\text{xz}}$\\
	-\frac{i}{2} & 0 & 0 & -\frac{i}{2} & \frac{\sqrt{3}}{2} & 0 & 0 & 0 & -\frac{1}{2} & 0 & $\quad d_{\text{xz}}$\\
	i & 0 & 0 & -\frac{i}{2} & 0 & 0 & 0 & -\frac{1}{2} & 0 & 0 & $\quad d_{x^2-y^2}$\\
	0 & -i & -\frac{i}{2} & 0 & 0 & 0 & \frac{1}{2} & 0 & 0 & 0 & $\quad d_{x^2-y^2}$\\
	\end{block}
	\end{blockarray}  \, \, \, .
	\end{equation*}
\end{widetext}

Please note that when we use the operators \eqref{eqn:R}, the order of the matrix elements are not the same as complex spherical harmonics and depends on the definition of the transformation matrix ($R_l$). In this work, the order of the real spherical harmonics are presented in Table \ref{tab1}.

\section{Iterative Minimization}
TBStudio algorithm is based on the Levenberg–Marquardt least-squares curve fitting approach in which we have a set of data obtained from first-principles calculation and an analytical expression representing the TB model for which we have to find the best independent parameters. Generally, the sum of the squares of the deviations may be written as follows

\begin{align}
S(\boldsymbol\beta) = \sum\limits_{n} \sum\limits_{\boldsymbol k} \sum\limits_{\sigma} (\varepsilon_{n \boldsymbol k \sigma}' - \varepsilon_{n \boldsymbol k \sigma}(\boldsymbol\beta))^2 \nonumber \, ,
\end{align}

\noindent where $\varepsilon$ and $\varepsilon'$ are, respectively, the TB results and DFT band energies and $\boldsymbol\beta$ is a parameter vector that is a set of independent variables i.e. SK parameters and overlap integrals and SOC. Non-linear least square minimization problems arise especially in curve fitting where here the curves are energy bands which are highly coupled to each other. To start the minimization, the user has to provide an initial guess for $\boldsymbol\beta$ and using the guessed $\boldsymbol\beta$ and the mentioned SK parameters one can find the TB band energies and also the Jacobian matrix

\begin{align}
J_{n \boldsymbol k \sigma,j} = \frac{\partial \varepsilon_{n \boldsymbol k \sigma}(\boldsymbol\beta)}{\partial \boldsymbol\beta_j} \nonumber \, .
\end{align}

In each iteration, we replace $\boldsymbol\beta$ by a new parameter vector $\boldsymbol\beta + \boldsymbol\delta$ in which $\boldsymbol\delta$ is a correction vector that can be estimated by first order Taylor series expansion of TB band energies as follows

\begin{align}
\varepsilon_{n \boldsymbol k \sigma}(\boldsymbol\beta + \boldsymbol\delta) = \varepsilon_{n \boldsymbol k \sigma}(\boldsymbol\beta) + {\boldsymbol J}_{n \boldsymbol k \sigma}.{\boldsymbol \delta} \nonumber \, .
\end{align}

At optimum values for the independent variables, the sum of square deviations has reached its minimum with respect to the independent vector and we have

\begin{align}
\frac{\partial S(\boldsymbol\beta + \boldsymbol\delta)}{\partial \boldsymbol\delta} = 0 \, ,
\end{align}

\noindent which however may also result in local minima. Depending on the problem and the system and the number of degrees of freedom, there might be many local minima which occur when the objective function value is greater than its value at the so-called global minimum. When multiple minima exist the important consequence is that the objective function definitely has a maximum value somewhere between two minima which makes it difficult to obtain a good fitting. Refinement from a bad SK parameter point (a set of independent parameter values $\boldsymbol\beta_0$) close to a physically unknown local minimum will be ill-conditioned and should be avoided as a starting point. Because generally we do not have an analytical expression for the TB band structure, we can not analyze the location of local minima in detail. Band structure curves have complex forms for which it is either very difficult or even impossible to derive analytical expressions for the elements of the Jacobian. In such cases, we need to find the elements by using numerical approximations as follows

\begin{align}
\frac{\partial \varepsilon_{n \boldsymbol k \sigma}(\boldsymbol\beta)}{\partial \boldsymbol\beta_j} \approx \frac{\varepsilon_{n \boldsymbol k \sigma}(\boldsymbol\beta+ \boldsymbol\delta) - \varepsilon_{n \boldsymbol k \sigma}(\boldsymbol\beta- \boldsymbol\delta)}{2\boldsymbol\delta_j} \nonumber \, .
\end{align}

\section{Explicit form for the Tight-Binding Hamiltonian}
\label{sec:exp}
\begin{figure}[b!]
	\includegraphics[width=0.9\linewidth]{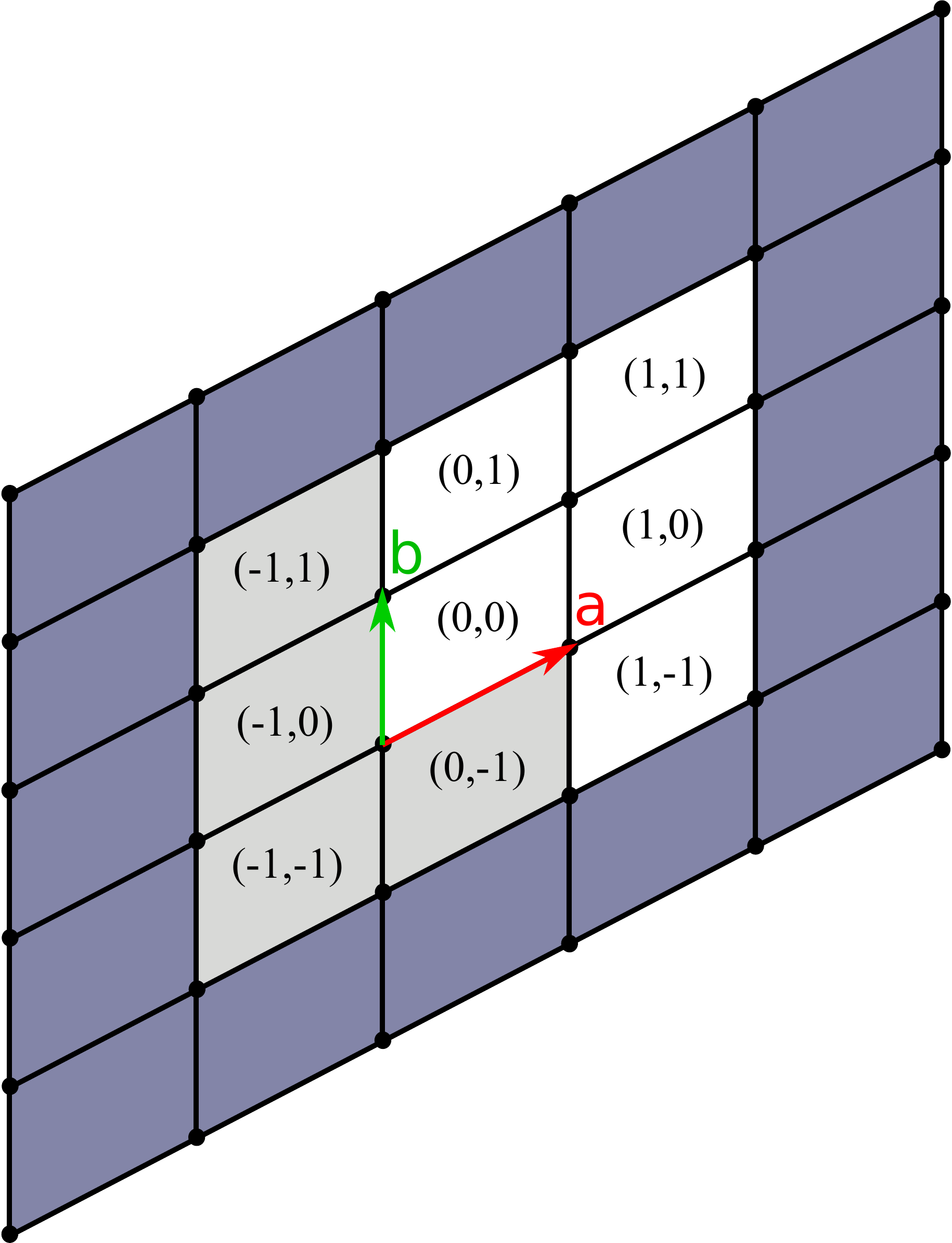}
	\caption{Schematic representation of the TB model for a typical 2D crystal. The red ($\textbf{a}$) and green ($\textbf{b}$) vectors are the unit vectors and the white cells indicate independent unit cells.}
	\label{fig:bl}
\end{figure}

\noindent Fig. \ref{fig:bl} shows a schematic representation of the TB model. The mono-electronic Hamiltonian $H$ and the overlap matrix $S$ may be rewritten as

\begin{align}
\centering
\hat{H} = \sum\limits_{n \in \mathbb{Z}} \sum\limits_{m \in \mathbb{Z}} \sum\limits_{l \in \mathbb{Z}} \hat{h}_{l,n,m} e^{i \mathbf{k}.\mathbf{R}_{l,m,n}} \, ,
\label{eqn:Hh}
\end{align}

\noindent and

\begin{align}
\centering
\hat{H} = \sum\limits_{n \in \mathbb{Z}} \sum\limits_{m \in \mathbb{Z}} \sum\limits_{l \in \mathbb{Z}} \hat{s}_{l,n,m} e^{i \mathbf{k}.\mathbf{R}_{l,m,n}} \, .
\label{eqn:Ss}
\end{align}

\noindent Since $\textbf{H}$ and $\textbf{S}$ are Hermitian, therefore

\begin{align}
\centering
\hat{h}_{-n,-m,-l} &= \hat{h}_{n,m,l}^\dagger \, , \nonumber \\
\hat{s}_{-n,-m,-l} &= \hat{s}_{n,m,l}^\dagger \, ,
\label{eqn:sh}
\end{align}

\noindent and thus, in this two dimensional example, we have only five independent matrices. As shown in Fig. \ref{fig:bl} we must determine only the matrices $\textbf{h}$ and $\textbf{s}$ for the cells at $(0,0)$, $(1,0)$, $(0,1)$, $(1,1)$ and $(-1,1)$. Using the SK coefficients we can calculate the Hamiltonian and the overlap matrix and extract the matrices $\textbf{h}$ and $\textbf{s}$. TBStudio generates the Hamiltonian and overlap matrix for any independent unit cell. Also the code generator tool builds Eqs. \eqref{eqn:Hh} and \eqref{eqn:Ss} in other desired programming languages. Also, one can use the outputs from TBStudio for post-processing in other transport packages. There are many useful packages for fast calculation of various physical properties of tight-binding models such as PyBinding \cite{cPyBinding}, Kite \cite{cKite}, Kwant \cite{cKwant}, GPUQT \cite{GPUQT}, TBTK \cite{TBTK}, PythTB \cite{PythTB} and WannierTools \cite{WannierTools}.

After determining the Hamiltonian and the overlap matrix, one can easily calculate the eigenstates and the Bloch functions as well using Eqs. \eqref{eqn:Bloch} and \eqref{eqn:lcao}. The $i^{th}$ Wannier function (WF) for the cell $(l,m,n)$ is the Fourier coefficient of $\psi_{i,k}(\textbf{r})$ as follows

\begin{align}
\centering
\textit{w}_{i}^{l,m,n}(r) = \frac{V}{8 \pi^3} \int {\boldsymbol d}^3 \boldsymbol{k} e^{-i \boldsymbol{k}.\boldsymbol{R}_{l,m,n}} \psi_{i,k}(\textbf{r}) \, .
\label{eqn:Ss}
\end{align}

The WF calculated by this method is very close to the real chemical bonding and provides a reliable insight into the nature of the orbitals in the study of electronic properties of solids. It should be noted that, in this way we do not have the problem which we encounter in finding WFs using Maximally Localized Wannier Functions (MLWF) method \cite{MLWF}. Practically, MLWF algorithms change the shape of WF mathematically to find a perfect fitting to regenerate the band-structure obtained by first-principle methods, because a set of WFs which can generate the obtained band-structure is not unique and may physically not a good set of atomic-orbital-like WFs. In such algorithms that has been implemented in Wannier90 \cite{wannier90} and OpenMX \cite{openmx}, to achieve a physically reliable set of WFs one needs to have a good initial guess and also follow the symmetry adapted methods, but it results in computational difficulties and convergence failure.

\section{Summary}
\label{Summary}
In summary, Tight-Binding Studio (TBStudio) is a user-friendly application in the field of quantum computing simulation. The significant parts and important abilities of the TBStudio were briefly mentioned. In short, using this software one can calculate the electron hopping between different orbitals of different atoms and generate an explicit Hamiltonian matrix to do post calculations using Green's function theory. TBStudio 
are the density functional theory and the Green's function theory in the tight-binding approximation. TBStudio is the first step of simulating electronic properties of solids and nanostructures (ie. dispersion, transmission, density of states, current, etc). Also, it is able to calculate thermodynamic properties by means of statistical mechanics approaches.
\section{Data Availability}
\label{DataAvailability}
The supporting information and several examples are available at \MYhref{https://tight-binding.com}{\textit{tight-binding.com}}. The examples and the supporting codes in additional programming languages, i.e. Matlab, Mathematica, Python, C, C++ and Fortran are also accessible through Code Generator tools in TBStudio.\\

\textbf{Acknowledgements} This work was supported by the Methusalem program of the Flemish government and M. Nakhaee was supported by a BOF-fellowship (UAntwerpen).


\end{document}